\documentstyle[prl,aps,multicol,epsf]{revtex}
\tighten
\narrowtext
\newcommand \be {\begin{equation}}
\newcommand \ee {\end{equation}}
\newcommand \bea {\begin{eqnarray}}
\newcommand \eea {\end{eqnarray}}
\begin{document}
\pagestyle{empty}
\draft
 \begin{multicols}{2}

{\bf \noindent Comment on ``Antilocalization in a 2D Electron
Gas in a Random Magnetic Field''} 

\vspace{3mm}

In a recent Letter \cite{taras}, Taras-Semchuk and Efetov reconsider the
problem of electron localization in a random magnetic field (RMF)
in two dimensions. They obtain an additional term in the effective
field theory ($\sigma$-model) of the problem, leading to
delocalization at the one-loop level. This calls into question the results 
of earlier analytical studies \cite{amw,network}, where the RMF
problem was mapped onto
the conventional unitary-class $\sigma$-model, implying that
the leading quantum correction is of two-loop order and of a
localizing nature.

We will show, however, that the new term that is claimed to appear in
\cite{taras} due to the long-range nature of the vector
potential (${\bf A}$) correlations, in fact
does not exist and was erroneously obtained in \cite{taras} because of
an inconsistent treatment violating gauge invariance. 
A diagrammatic analysis of the effective action (9) of
Ref.~\cite{taras} shows that in 
leading order the contribution of the new term to the conductivity 
(denoted there as $\langle F_\parallel^2\rangle$) is represented by a
diagram with one diffuson ($m=0$ angular harmonic in
momentum space) and two ``massive diffusons'' ($m\ge 1$), see Fig. 1a. In
contrast to the 
normal ($m=0$) diffuson, for which summation of an infinite series of
ladder diagrams produces a singularity at $q,\omega\to 0$, the
corresponding terms of the series for the ``massive diffuson'' can be
considered separately.  
In particular, keeping the first term in both  ``massive diffusons''
and switching to the conventional diagram technique, we get the diagram
of Fig. 1b; higher terms produce analogous diagrams with $n>2$
impurity lines crossing the diffuson. 

\begin{figure}
\narrowtext
\centerline{ {\epsfysize=2cm{\epsfbox{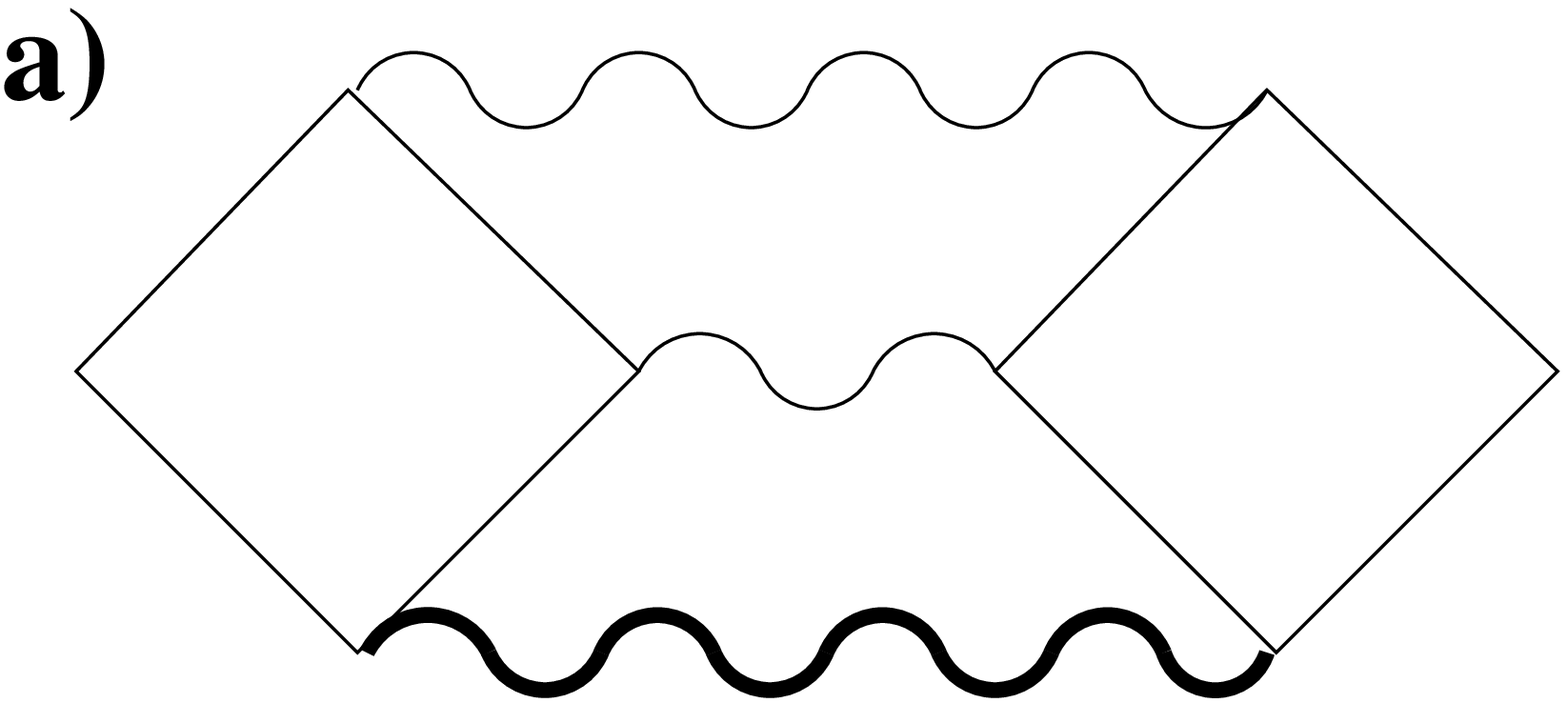}}} \hspace{4mm}
{\epsfysize=2cm{\epsfbox{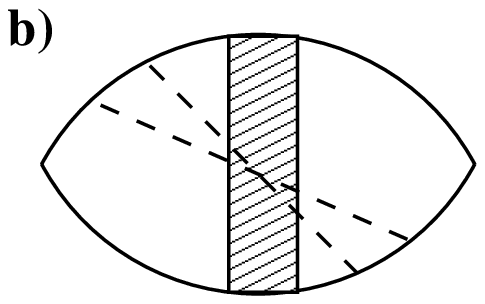}}} }
\vspace{3mm}
\caption{a) Leading diagram for $\langle F_\parallel^2\rangle$
 with one normal diffuson (thick wavy line) and two high-$m$
 ``diffusons'' (thin wavy lines); b) corresponding contribution in the
 impurity diagram technique; the shaded block denotes the diffuson.}
\end{figure}

We show now that for any  $n$ the sum of {\it all}  diagrams with
$n$ impurity lines crossing the diffuson and having their starting
(end) point anywhere on the left (right) block yields a non-divergent
contribution to the conductivity. Indeed, this contribution has the form
\begin{eqnarray}
\Delta\sigma_{xx} & = & {e^2\over2\pi}{1\over n!}\int (dq_1)\ldots
(dq_n) {1\over 2\pi\nu\tau^2(Dq^2-i\omega)} \nonumber \\
& \times & S_{x\alpha_1\ldots\alpha_n}({\bf q}_1,\ldots,{\bf q}_n)
 S_{x\beta_1\ldots\beta_n}({\bf q}_1,\ldots,{\bf q}_n)\nonumber \\
& \times & {\cal D}_{\alpha_1\beta_1}({\bf q}_1)\ldots
{\cal D}_{\alpha_n\beta_n}({\bf q}_n), \label{e1}
\end{eqnarray}
where ${\bf q}=\sum_i{\bf q}_i$ is the diffuson momentum, 
${\cal D}_{\alpha\beta}(q)=[W_B(q)/q^2]
(\delta_{\alpha\beta}-q_\alpha q_\beta/q^2)$, and 
$W_B(q)=\langle B(q)B(-q)\rangle$  is the RMF correlation function in
momentum space. Exploiting the fact that insertion of a RMF line
in all possible ways can be generated by variation of the vertex part $S$
with respect to the vector potential, 
$\delta/\delta A_\alpha({\bf r})$, and using the gauge invariance
argument, it is not difficult to show that the vertex part satisfies 
$$
S_{x\alpha_1\ldots\alpha_n}({\bf q}_1,\ldots,{\bf q}_n)\propto
qq_1\ldots q_n\ .
$$
Therefore, all the singularities in the vector potential correlators
${\cal D}_{\alpha_i\beta_i}$
and in the diffusion propagator in Eq.~(\ref{e1}) are canceled by the
vertex parts, yielding a non-divergent contribution of the form
$$
\Delta\sigma_{xx}\propto \int (dq)q^2/(Dq^2-i\omega) \propto {\rm
const} + |\omega|\tau\ .
$$
Though this correction is non-analytic at $\omega\to 0$ (corresponding
to a $1/t^2$ long-time tail in the velocity correlation function, see
\cite{wilke} for details), it is non-divergent. For a random
potential, a similar cancellation of the divergent diffuson
contribution occurs \cite{vw}.

Having shown that a divergent one-diffuson contribution does not
exist, it is natural to ask why did the authors of \cite{taras} find
it. The point is that while the sum of all diagrams with a given $n$
is finite, any individual diagram does diverge. So, a loss of some
diagrams produces a spurious divergent term. It remains to
point out a diagram lost in \cite{taras}. In fact, there are many of
them: all the diagrams originating from the ${\bf A}^2$ term in the
Hamiltonian are omitted. While these diagrams do not contribute in the
derivation of the diffusion action in leading order performed in
\cite{amw}, they are crucially important for maintaining gauge
invariance in higher order calculations.

This work was supported by SFB 195 and the Schwerpunktprogramm
``Quanten-Hall-Systeme'' of the Deutsche Forschungsgemeinschaft.

\vspace{5mm}

\noindent Alexander D. Mirlin and Peter W\"olfle\\
{\small \indent Institut f\"ur Nanotechnologie,\\
        \indent Forschungszentrum Karlsruhe,
        76021 Karlsruhe, Germany;\\ [1mm]
        \indent Institut f\"ur Theorie der kondensierten Materie,\\
         \indent Universit\"at Karlsruhe,
         76128 Karlsruhe, Germany}

\pacs{\hspace{-2cm} PACS numbers: 73.20.Fz, 72.15.Rn}

\end{multicols}

\end{document}